\documentclass[10pt,letterpaper]{article}
\usepackage{opex3}
\usepackage{epsfig}

\newcommand{\mum}{\ensuremath{\mu \mathrm{m} }}

\newcommand{\Htwo}{\ensuremath{\mathrm H_{2}}}

\makeatletter
\def\blfootnote{\xdef\@thefnmark{}\@footnotetext}
\makeatother

\begin{document}


\title{Multipass laser cavity for efficient transverse
  illumination of an elongated volume}

\author{
Jan Vogelsang,$^1$
Marc Diepold,$^1$
Aldo Antognini,$^2$
Andreas Dax,$^3$
Johannes G\"otzfried,$^1$
Theodor W.\ H{\"a}nsch,$^1$
Franz Kottmann,$^2$
Julian J. Krauth,$^1$
Yi-Wei Liu,$^4$
Tobias Nebel,$^1$
Francois Nez,$^5$
Karsten Schuhmann,$^{2,3}$
David Taqqu,$^3$
and Randolf Pohl$^{1,*}$}

\address{
$^1$Max--Planck--Institut f{\"u}r Quantenoptik, 85748 Garching, Germany\\
$^2$Institute for Particle Physics, ETH Zurich, 8093 Zurich, Switzerland\\
$^3$Paul Scherrer Institute, 5232 Villigen--PSI, Switzerland\\
$^4$Physics Department, National Tsing Hua University, Hsinchu 300, Taiwan\\
$^5$Laboratoire Kastler Brossel, \'Ecole Normale Sup\'erieure, CNRS\\and Universit\'e P.~et M.~Curie, 75252 Paris, CEDEX 05, France\\}

\email{$^*$randolf.pohl@mpq.mpg.de} 



\begin{abstract}
  A multipass laser cavity is presented which can be used to illuminate an
  elongated volume from a transverse direction. The illuminated volume can
  also have a very large transverse cross section.  Convenient access to the
  illuminated volume is granted.
  The multipass cavity is
  very robust against misalignment, and no active stabilization is needed.
  The scheme is suitable for example in beam experiments, where the beam path
  must not be blocked by a laser mirror, or if the illuminated volume must be
  very large.
  This cavity was used for the muonic-hydrogen experiment in which 
  6\,\mum{} laser light illuminated a volume of 
  $7 \times 25 \times 176$\,mm$^3$, using mirrors that are only 12\,mm 
  in height.
  We present our measurement of the intensity distribution inside the
  multipass cavity and show that this is in good agreement with our
  simulation.
\end{abstract}

\ocis{(080.4035) Mirror system design.}
%
%
%
%
%



\section{Introduction}

Experiments with particle beams at times require illumination of an elongated
volume, i.e.\ over a length which is much larger than the transverse dimension
of the beam.  One solution would be to intersect a laser beam at some finite
angle with respect to the particle beam such that the particle beam is not
obscured by laser mirrors. This usually results in a poor overlap between the
particle beam and the laser beam.

In addition, such an arrangement allows only illumination of a volume with
small transverse dimensions if a cavity is used to increase the laser
fluence.  Cavities with larger beam waists would require cavity mirrors with
exceedingly large radii of curvature, which makes such a cavity unstable.

Here we present a multipass mirror cavity that can be used to efficiently
illuminate, from the transverse direction, a volume with a length of several 
tens of cm or more, limited only by the mirror reflectivity. In addition, 
the illuminated volume can have a large transverse size.
The beam volume is not obscured by the mirrors, allowing for
large-solid-angle detection of photons. In addition, the mirror cavity is
extremely robust against misalignments making control and stabilization of the
mirrors unnecessary.

\begin{figure}[b]
\centering\includegraphics[width=0.9\textwidth]{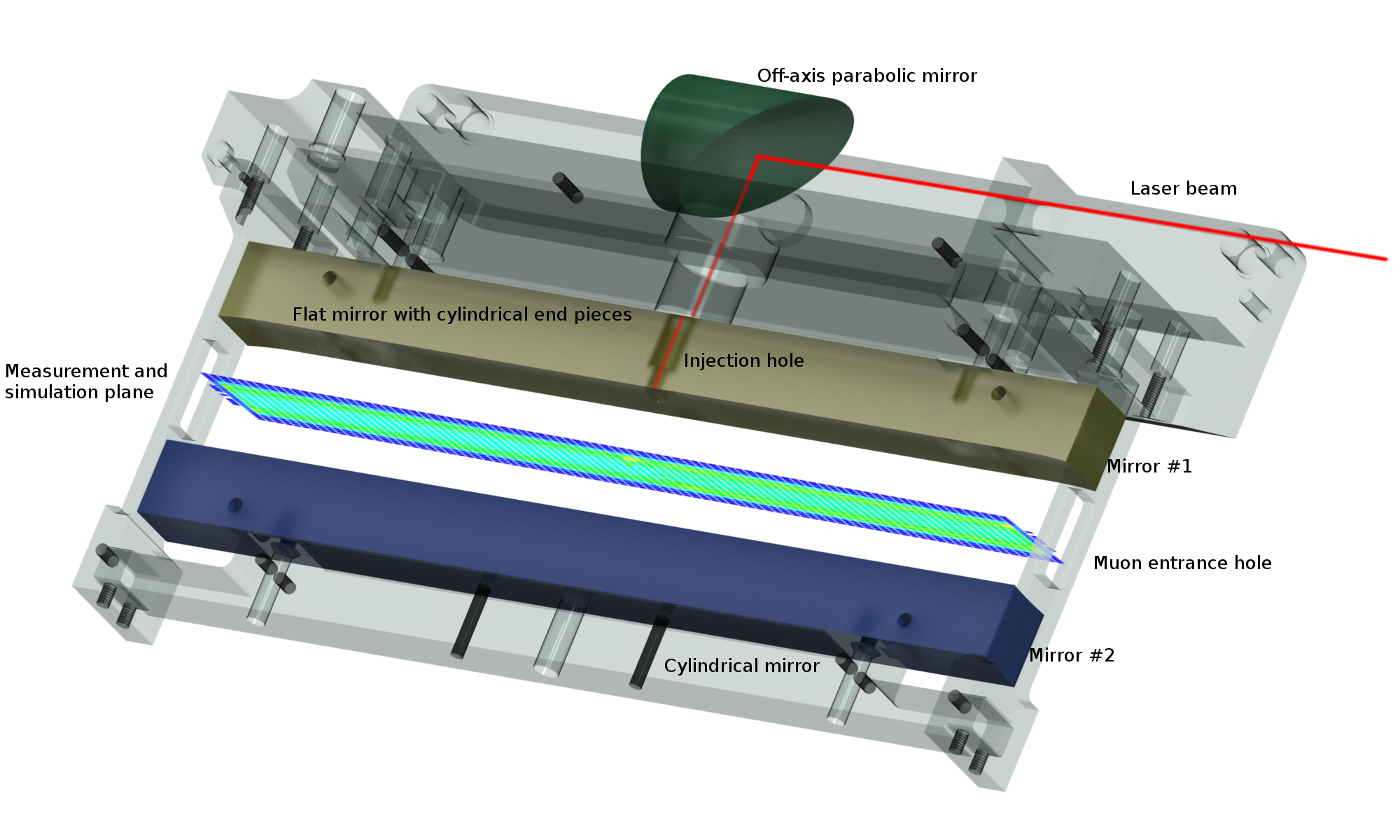}
\caption{Complete setup of the multipass laser cavity, with mounts and 
  injection optics. Also indicated is the cavity mid-plane where the 
  intensity distribution is measured and simulated.
}
\label{fig:cavity}
\end{figure}

The mirror cavity shown in Fig.~\ref{fig:cavity} was used to illuminate the
200\,mm long stop volume of muons injected into low pressure \Htwo{} gas. The
setup was used for the recent determination of the proton charge radius via
laser spectroscopy of the exotic muonic hydrogen
atom~\cite{Pohl:2010:Nature_mup1,Antognini:2013:Science_mup2}.
The low gas pressure of only 1\,mbar (at room temperature) is
necessary to avoid collisional processes which shorten the lifetime of
the muonic hydrogen atoms~\cite{Pohl:2006:MupLL2S}. The low gas
density results in an approximately pencil-sized stop volume with a
length of 200\,mm.
The transverse dimension of the muon stop volume to be illuminated is
12\,mm (width) times 5\,mm (height).
Several large-area avalanche photo diodes~\cite{Fernandes:2003:NIM,Ludhova:2005:LAAPDs}
 ($14 \times 14$\,mm$^2$ active
surface) were placed as close as 8\,mm from the muon beam axis to
maximize the solid angle for the detection of x-rays from muonic hydrogen.

Each incoming muon triggers a pulsed laser
system~\cite{Antognini:2005:6mumLaser,Antognini:2009:Disklaser},
tunable around $\lambda \approx 6\,\mum{}$. We obtained pulses of
0.25\,mJ energy, with a pulse length of 5\,ns.
The multipass mirror cavity described in here was used to illuminate
the elongated muon stop volume from the transverse direction. We
realized a rather homogeneous $(\pm 30\%$) illumination of the
stop volume and confined the laser pulse for approximately 55\,ns,
corresponding to about 670 reflections of the light inside the cavity,
limited by the reflectivity of the coating of
$\rm R \approx 99.89\,\%$.

The cavity is installed inside a superconducting 5\,Tesla solenoid
which is then evacuated. Our cavity design proved to be very robust
against misalignments, and no active stabilization was required over
weeks of continuous data taking.

Three different cavities have been built with their main difference
being the high reflective (HR) coating (see Sec.~\ref{sec:coating}):
\begin{itemize}
\item{Cavity~A had a ThF$_4$ / ZnSe HR coating optimized for
  $\lambda \in [ 5.5 \dots 6.1]$\,\mum{}, but displayed a
  reflectivity better than 99.9\,\% around 850\,nm. This cavity was used in a
  test setup to produce most of the plots shown here, using a laser
  diode at 850\,nm.}
\item{Cavity~B had a Ge/ZnS coating, again optimized for
  $\lambda \in [ 5.5 \dots 6.1]$\,\mum{}. It was used for the muonic
  hydrogen measurements reported in
  Refs.~\cite{Pohl:2010:Nature_mup1,Antognini:2013:Science_mup2} and
  displayed the abovementioned light confinement time of 55\,ns 
  (Fig.~\ref{fig:cavity_decay_time}).}
\item{Cavity~C was recently used in a follow-up experiment for
  the laser spectroscopy of muonic helium ions at a wavelength
  between 800\,nm and 820\,nm~\cite{Antognini:2011:Conf:PSAS2010}.
  Here, a light confinement time of $\sim 100$\,ns was routinely
  observed for pulse energies reaching 4\,mJ.}
\end{itemize}

In the next section we discuss the geometry of the multipass cavity as
dictated by the experimental constraints. 
We then discuss the peculiarities of the light injection, present intensity
distributions measured for various misalignments and compare them to
simulations.

\begin{figure}[t]
\centering\includegraphics[width=0.8\textwidth]{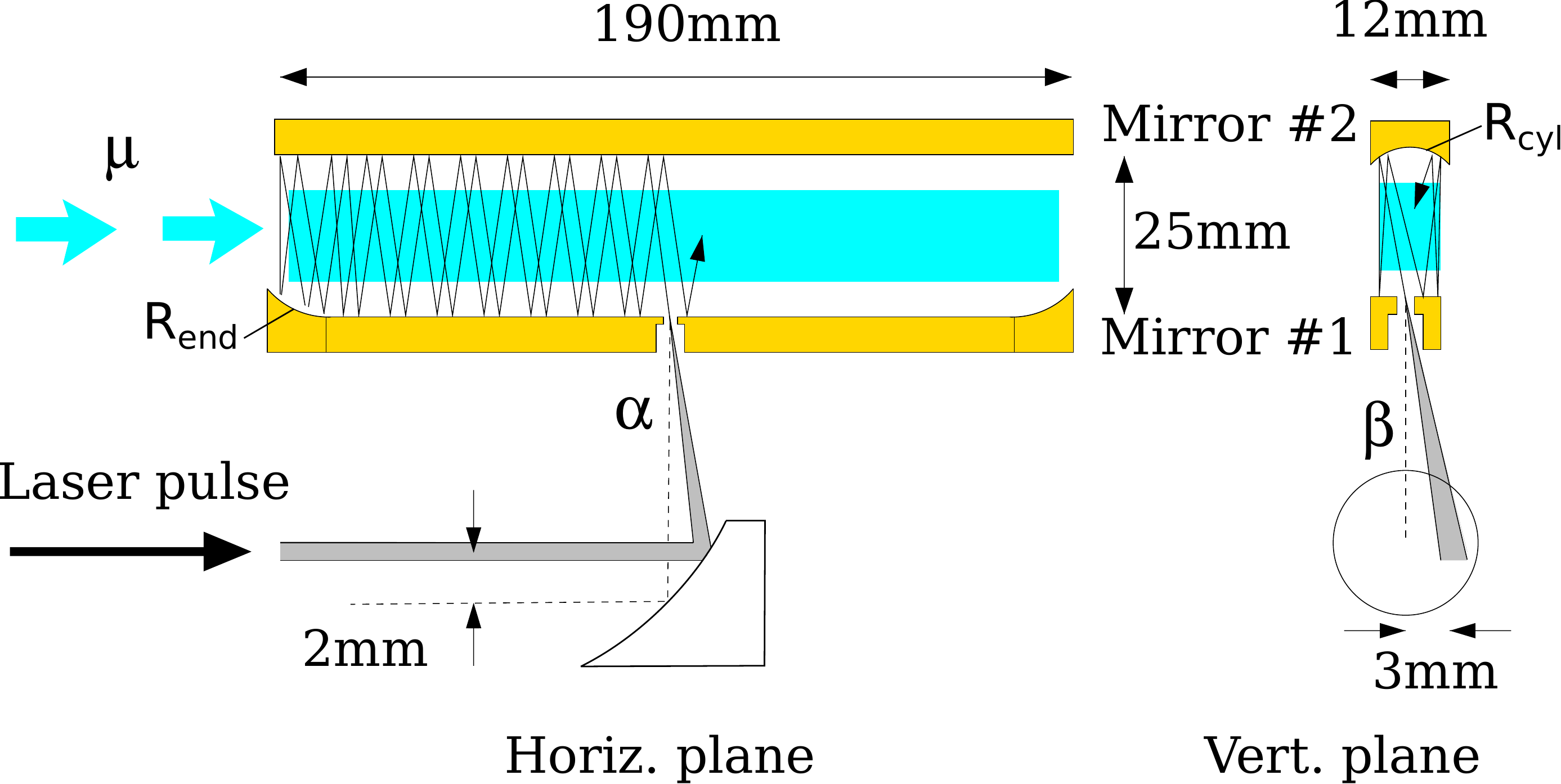}
\caption{Sketch of the cavity, indicating the light incoupling via the
  off-axis parabolic mirror (OAP).  The blue shaded region indicates
  the muon stop volume to be illuminated by the light.  The light is
  injected at an angle (horizontal $\alpha$=40\,mrad, vertical
  $\beta$=65\,mrad) by a (parallel) displacement of the incoming light
  by 2 and 3.3\,mm, respectively.  The OAP has an off-axis focal length
  of 50.8\,mm.}
\label{fig:incoupling}
\end{figure}

\section{Geometry}

The multipass cavity consists of two mirrors, each 190\,mm long and 12\,mm
high, and separated by 25\,mm, 
as shown in Figs.~\ref{fig:cavity} and \ref{fig:incoupling}.
The mirror length was limited to 190\,mm by geometrical constraints in our
setup even though the stop volume has a length of 200\,mm. The illuminated
volume was approximately 176\,mm long.
The thickness of the fused silica substrates was chosen to be 15\,mm.

Mirror \#1 (olive in Fig.~\ref{fig:cavity}) is assembled from a
170\,mm long flat central piece and two 10\,mm long cylindrical end
pieces that ensure horizontal (along the muon beam axis) confinement
of the light.
The cylindrical end pieces were bolted to the flat center piece as detailed in
Sec.~\ref{sec:practical}.

According to simulations of the cavity (see
Sec.~\ref{sec:simulation}), the radius of curvature of the cylindrical
end pieces, $R_{\rm end}$, has to be twice or four times the mirror
spacing to ensure efficient light confinement. We chose a mirror
spacing of 25\,mm and hence an end piece radius of curvature of
$R_{\rm end} = 100$\,mm.

Mirror \#2 (blue in Fig.~\ref{fig:cavity}) is a cylindrical mirror
(cylinder axis along the muon beam axis) that ensures the vertical
confinement of the light. Its radius of curvature $R_{\rm cyl}$ must {\em
  not} be twice or four times the mirror distance to avoid the laser
beam retracing itself after a few round trips. This would result in a
regular laser spot pattern instead of a washed-out homogeneous intensity
distribution.  We chose $R_{\rm cyl} = 110$\,mm.

\section{Light injection}
\label{sec:injection}

Light is coupled into the multipass cavity through a hole in the
middle of the flat piece of mirror~\#1. The hole has a diameter of
only 0.63\,mm at the optical surface of the mirror, and increases to
6\,mm diameter on the back side of the mirror. Instead of a conical
hole, we chose to drill three concentric holes (0.63\,mm $\times$
1\,mm length, 3\,mm diameter $\times$ 4\,mm, and 6\,mm diameter
$\times$ 10\,mm). The 0.63\,mm diameter hole was drilled using
ultrasonic drilling~\cite{Company:Dama}.

The laser light is focused into the center of the 1\,mm long smallest part
(0.63\,mm diameter) of the injection hole, with a waist of 100\,\mum{}.  For
the geometry presented here, a lens with a focal length of f=50.8\,mm could be
used for our laser beam that has a wavelength of 6\,\mum{} and a 1\,mm waist.
We, however, chose to use a 90$^\circ$ off-axis parabolic mirror (OAP), due to
geometric constraints inside the bore hole of our superconducting magnet.

The focusing element (lens or OAP) is used to control the size of the
illuminated volume inside the cavity, as shown in Fig.~\ref{fig:incoupling}:
A parallel displacement of the incoming laser beam causes the light to
enter the cavity at an angle ($\alpha, \beta$ in
Fig.~\ref{fig:incoupling}) with respect to the optical axis.  A vertical
displacement of 3.3\,mm with respect to the OAP center results in a
vertical injection angle of $\beta$=65\,mrad and consequently a
vertical size of the illuminated volume of 7\,mm. Note that the
cylindrical mirror \#2 ensures that the light remains confined up to a
maximum injection angle of 90\,mrad.

Similarly, a horizontal laser light displacement of 2\,mm on the OAP results
in a horizontal injection angle of $\alpha$=40\,mrad. The cylindrical end
pieces of mirror \#1 ensure horizontal confinement of the light and a length
of the illuminated volume of 176\,mm.

A convenient side effect of the OAP focusing is that small beam pointing
instabilities of our laser, which is $\sim 20$\,m away from the OAP, will only
result in parallel displacements of the laser beam on the OAP. The OAP will
still ensure that all light will enter the cavity through the injection
hole. Only the height of the illuminated volume may be slightly affected by
laser beam pointing instabilities.

Light leaving the cavity through the injection hole 
is efficiently collected by the OAP.  Monitoring this light reveals the light
storage time and hence the quality of the light confinement inside the cavity.
If the injection angle is chosen too small ($\alpha, \beta \approx 0$), 
most of the light escapes after one reflection. 
Efficient confinement is ensured only
for $\alpha, \beta $ larger than three 1/e angles of the light focused into
the injection hole.

\section{Intensity distribution}

We proceed now explaining how the light fills the cavity. We show simulations
performed using simple ray optics as detailed in Sec.~\ref{sec:simulation},
and measurements using light reflected from a 25\,\mum{} thin wire that is
moved through the center of the cavity, as explained in 
Sec.~\ref{sec:intens_meas}.

\subsection{The first roundtrips}
\label{sec:filling_cavity}

\begin{figure}[t]
\centering\includegraphics[width=0.99\textwidth]{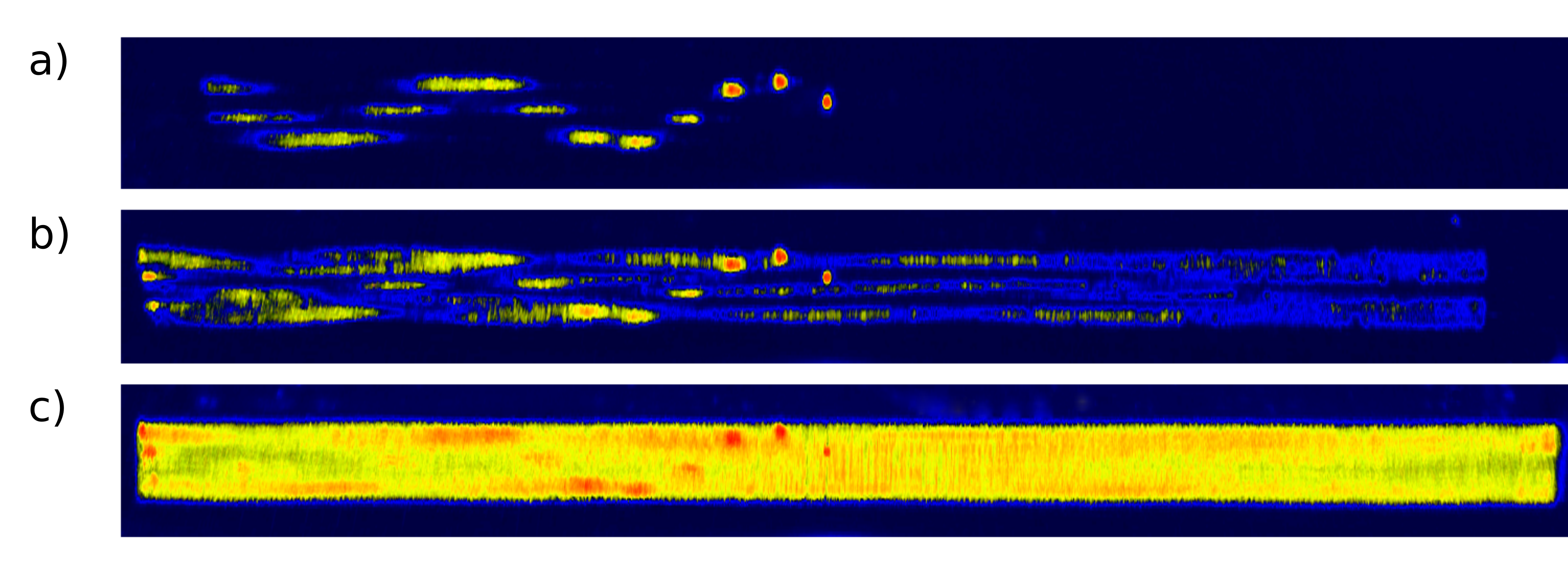}
\caption{Three intensity distributions measured in the cavity midplane
  (see Fig.~\ref{fig:cavity}). The distribution is measured at 850\,nm 
  in Cavity~A using
  reflections from a thin moving wire inside the cavity (see
  Sec.~\ref{sec:intens_meas} for details). Injected light started
  towards the top left. a: An absorber was placed in the left part of
  the cavity making the first quarter of the light roundtrip
  visible. b: The absorber was moved to the right end of the cavity.
  3/4 of a round trip is seen. c: No absorber. The light fills the
  cavity volume.  }
\label{fig:3steps}
\end{figure}

Figure~\ref{fig:3steps} illustrates the working principle of the
cavity. Shown are three measurements of the intensity pattern in the
cavity midplane (Fig.~\ref{fig:cavity}).

Light injected at angles $\alpha, \beta$ (see
Fig.~\ref{fig:incoupling}) travels first to the left. The cylindrical
mirror \#2 ensures vertical confinement of the light. This gives rise
to the vertical oscillation of the laser spot inside the cavity as it
travels to the left.  In Fig.~\ref{fig:3steps}(a) the light is
intentionally absorbed before it reaches the cylindrical end piece of
mirror \#1. One can also observe how the initially focused beam
spreads out in the horizontal direction.

In Fig.~\ref{fig:3steps}(b) the absorber was placed on the right hand
side of the cavity. The left cylindrical end piece of mirror \#1
causes the light to ``turn around'' and travel all the way to the
right end of the cavity.

Figure~\ref{fig:3steps}(c) shows the intensity distribution measured
inside the cavity without any absorber present. The volume is
filled homogeneously. The rather low reflectivity of the
mirrors used in this measurement prevents an even more homogeneous
intensity distribution. In addition, the measurement procedure is at
least partially responsible for the hot spots of the first few
reflections. The light hits the wire used for measuring the
distribution directly after entering the cavity. Part of the light is
scattered out of the cavity and is no longer available for filling the
space above and below the hot spot after one or more round trips in
the cavity.

\subsection{Homogeneously filled cavity}
\label{sec:intens_general}

\begin{figure}
\centering\includegraphics[width=0.99\textwidth]{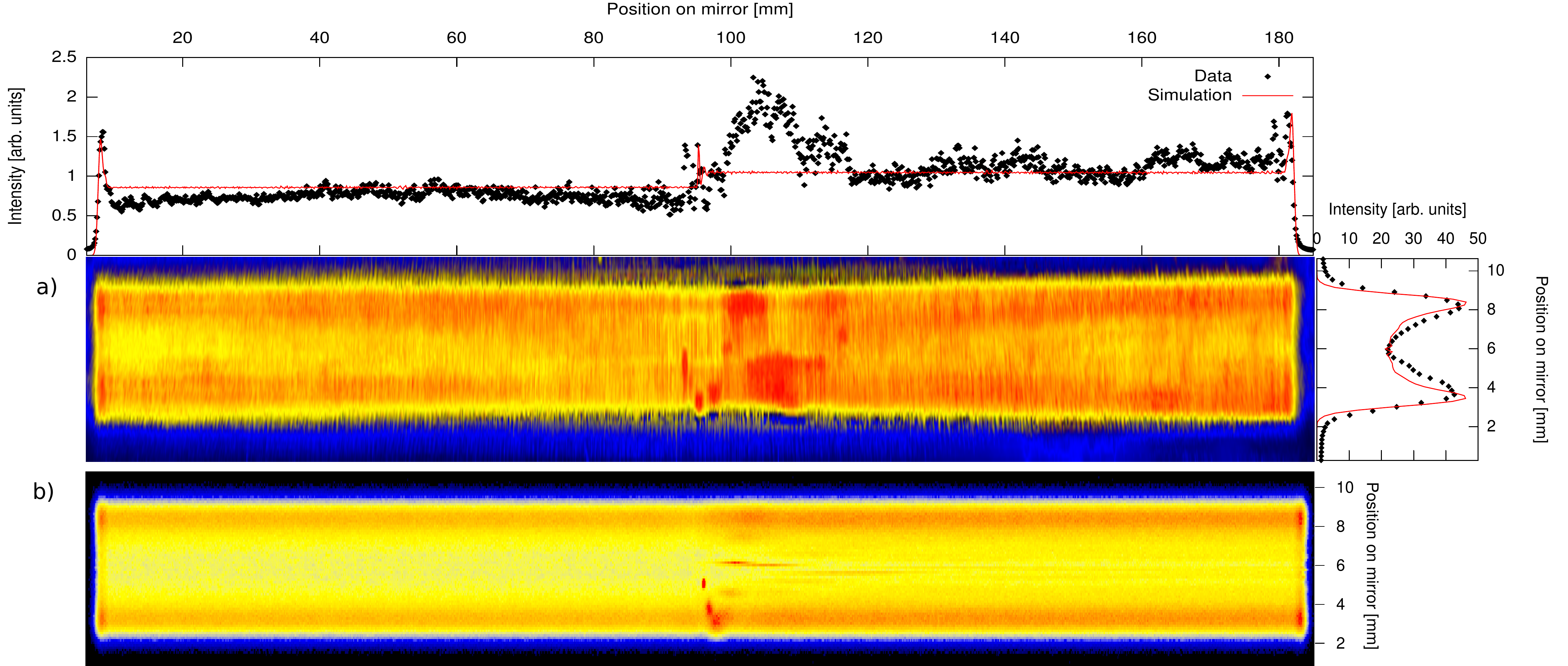}
\caption{The optimal distribution of light inside the cavity (a),
  compared to simulations (b, more details in
  Sec.~\ref{sec:simulation})).  Top: Intensity along the long
  axis. The spike in the red curve at z=95\,mm indicates the location 
  of the entrance hole.
  Right: Intensity along the vertical axis. For
  details see Sec.~\ref{sec:intens_general}. Light injection is
  towards the bottom right.}
\label{fig:flat}
\end{figure}

Figure~\ref{fig:flat} shows the intensity distribution measured in the mid
plane (see Fig.~\ref{fig:cavity}) of a well-aligned cavity.
Also shown are projections of the intensity in the horizontal and
vertical direction, together with simulations (see
Sec.~\ref{sec:simulation}). The agreement is rather good, except for
the region just right of the injection hole, where the data is well
above the simulation.  This is, however, believed to be an artifact of
the measurement process as explained in Sec.~\ref{sec:intens_meas}.

Several features are visible in the projections. The step at the center of the
horizontal distribution originates from the horizontal injection angle
$\alpha$ (see Fig.~\ref{fig:incoupling}):
The light will first travel to the right, and several loss mechanisms reduce
the intensity before the light arrives in the left part of the cavity:
finite mirror reflectivity,
possible imperfections at the transition from the 
flat to the cylindrical end piece of mirror \#1, 
and the losses at the injection hole after half a roundtrip.

The higher intensity at both ends of the horizontal distribution can be
understood in a ray optics picture: A ray arriving at the cylindrical end
piece with an angle $\alpha$ will gradually reduce its angle in the
cylindrical region until it eventually ``turns around''. This however results
in an increased intensity, as the ray undergoes more reflections (per mm) at
the turning point.

The same effect leads to the double-peak structure observed in the vertical
intensity distribution: A ray initially traveling downwards at an angle $\beta$
will gradually reduce its angle, perform more reflections per unit distance
close to the lower turning point, but quickly traverse the central part of the
mirror. 
This has the beneficial side-effect that the losses through the injection hole
are smaller than expected from the average fluence inside the illuminated
volume.

\begin{figure}[t]
\centering\includegraphics[width=0.99\textwidth]{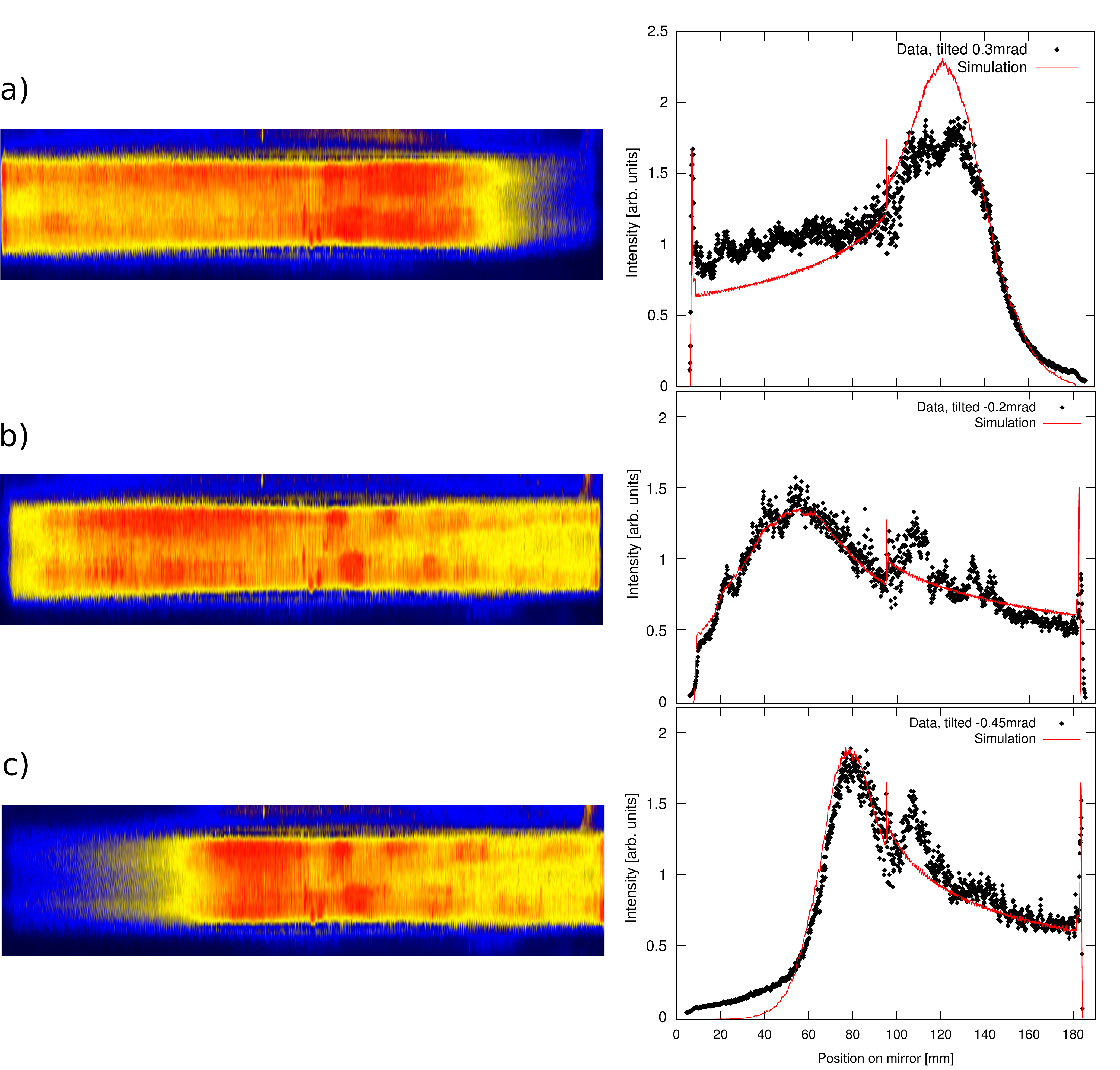}
\caption{Light distribution inside the cavity for different
  misalignments of mirror \#1 around the vertical axis. The right
  images show the mean intensity of the left images in horizontal
  direction. Such a strong misalignment has never been observed in the
  experiment. However, the light is still confined in the cavity, even
  though mostly on one side.}
\label{fig:tilted_aug}
\end{figure}

\subsection{Mirror misalignments}

Robustness against mirror misalignments was one of the design goals of
the cavity. In our
experiment~\cite{Pohl:2010:Nature_mup1,Antognini:2013:Science_mup2} we
aligned the cavity mirrors and the injection optics (OAP) by measuring
the storage time of the light (6\mum{} wavelength) inside the
cavity. The alignment procedure consists of aligning the mirrors by
eye and final optimization of the storage time by tilting the
mirrors. The first alignment does not have to be very accurate since
the light with a wide range of angles is confined between the mirrors.

We closed and evacuated the setup and switched on a high magnetic
field of 5\,T. No remote-controlled or active mirror stabilization was
employed. Still, no mirror misalignment was observed over the course
of a few weeks.

This robustness originates from the small curvature radii of the employed
cylinders: The 10\,mm long cylindrical end pieces of mirror \#1
with $R_{\rm end} = 100$\,mm can capture light up to (horizontal) angles
$\alpha$ of roughly 160\,mrad.
This corresponds to maximally allowed mirror misalignments of about $\pm
3$\,mrad (around the vertical axis), as the rays accumulate twice the mirror
misalignment on each reflection of the misaligned mirror. For such enormous
misalignments the light will only illuminate half of the enclosed volume,
though.

Measurements of very misaligned cavity mirrors can be found in
Fig.~\ref{fig:tilted_aug}, together with simulations that describe the data
quite well.
Here, we tilted one of the mirrors by up to 0.45\,mrad, corresponding to
85\,\mum{} displacement over the mirror length of 190\,mm. Still, three
quarters of the cavity volume are filled with light, and no light is lost.

Vertical misalignments, i.e.\ rotation around the long axes of the mirrors,
are much less critical. Such a mirror tilt is equivalent to a vertical
displacement of the cylindrical mirror \#2.
The effect is hence only a vertical displacement of the illuminated
volume.  The strongly focusing cylindrical mirror ensures light confinement
for vertical mirror tilts up to 10\,mrad in the geometry presented here.

\section{Simulation and measurement}
\label{sec:intensity}

\subsection{Simulation}
\label{sec:simulation}

The intensity distribution was simulated using a simple ray-optics
simulation. The simulation included the full geometry, including the
possibility of misaligned mirrors. 

It turns out that, for the curvature radii presented here, 
it is not necessary to take into account the exact
reflection point in 3 dimensions between a ray and the curved and
possibly tilted mirror surfaces. 
The cavity presented here could as well have been simulated by calculating the
intersection of rays with flat mirrors and taking the curvature into
account only for the reflection angle.  Also, a simulation of Gaussian
beams in the vertical dimension, propagated using ABCD-matrices, did
not produce very different results.

The simulation used about $10^4$ rays. Their initial position and angle in the
injection hole were distributed according to the assumed Gaussian TEM$_{00}$
mode of the injected laser beam. A simulation takes about a minute on a typical
contemporary PC.

The simulation created the intensity profile on various planes parallel to the
mirrors, as well as the time spectrum of the light leaving the cavity through
the injection hole. This light was also used in the experiment to monitor
adequate light confinement.

Interference effects are not expected to be important. The injected light is
strongly focused, giving a large variety of ray angles. After a few
roundtrips, every point receives light from many different directions.

\subsection{Measurement of the intensity distribution in the cavity midplane}
\label{sec:intens_meas}

Cavity~A was used for measuring the light distribution between the
mirrors. A laser diode emitting light at 850\,nm was used for these
measurements. Although the reflective coating is optimized for a
wavelength of 6\mum{}, it has a sufficiently high reflectivity of
more than 99.9\% around 850\,nm. 
Simulations showed that the
distribution inside the cavity can be well estimated using ray optics
with only minor influence of the laser wavelength.

The intensity distribution inside the cavity is measured by recording
the light scattered from a thin wire (25\mum{} diameter) which is
moved along the long axis of the cavity mirrors. A video of the wire
moving at constant velocity is recorded. The video is then converted
into a single image by taking the maximum value recorded per pixel.

In general, the 25\mum{} diameter of the wire is small enough to not severeley
affect the light confinement. As mentioned above, we don't expect intensity
variations on very small length scales, as interference is not expected to
play a relevant role.

The finite diameter of the wire does however have an influence on measurement 
of the light distribution, causing artifacts like
an overestimated intensity of the injected light and an apparent
constriction of the illuminated volume next to the injection hole (see
Fig.~\ref{fig:flat}).
Here, the injected light is still well collimated, and the wire can scatter a
large fraction of the light. 

\subsection{Online monitoring of the intensity distribution}
\label{sec:intens_mon}

Cavity~C used to measure several transition
frequencies in muonic helium ions~\cite{Antognini:2011:Conf:PSAS2010}
at wavelengths $\lambda \in [800 \dots 960]$\,nm.  Here, photo diodes
(PD) have been used for online monitoring of the intensity
distribution. Each PD is connected to a $\sim 10$\,mm long
optical multi-mode fiber with a core diameter of a few hundred \mum{}.
Six of these PD-fiber-assemblies have been mounted in the support of
the cylindrical mirror \#2, and detect the light transmitted through
the HR coating.

Four fibers are mounted close to the corners of the desired light distribution
(i.e.\ at y=$\pm 3.5$\,mm in height, and z=$\pm 80$\,mm along the beam axis).
The intensity distribution can be easily optimized by equalizing and
maximizing the signal on all 4 PDs simultaneously. This enables us to optimize
both the cavity mirror tilts and the light injection angle within a few
minutes.

Two additional PD-fiber-assemblies monitor the light on the axis of the cavity
(i.e\ at height y=0) where most of the atoms are.

The multi-mode fiber is used to limit both the light intensity experienced by
the PDs and the field-of-view of each detector, thereby increasing the
sensitivity to misalignments.

\subsection{Fluence estimate and light decay time}
\label{sec:fluence}

The full simulation agrees very well with the measured intensity distribution
and the lifetime of the light escaping through the injection hole.

However, already a simple estimate gives the average laser fluence achieved 
by the
mirror setup: The laser light at $\lambda = 6$\,\mum{} illuminates a
region with an area of $176 \times 7$\,mm$^2$.

The losses are as follows:
The mirror reflectivity or $R=99.89$\,\% results in $L_{\rm refl} = 11 \times
10^{-4}$ per reflection.
The injection hole with a diameter of 0.63\,mm creates losses of 
$L_{\rm  hole} \approx
2/3 \times \pi/4 \times (0.63\,{\rm mm})^2 / (7 \times 176 {\rm mm}^2) 
= 1.7 \times 10^{-4}$ 
every {\em second} reflection (i.e.\ for each reflection on mirror \#1). 
The additional factor of $\sim 2/3$ arises from the reduced laser fluence at
the entrance hole location (cavity axis), see Fig.~\ref{fig:flat} and
Sec.~\ref{sec:intens_general}.

The two gaps of $\sim 50$\,\mum{} width between the flat and the
cylindrical parts of mirror \#1 result in losses of $L_{\rm gap} = (2 \times
0.05\,{\rm mm}) / 176\,{\rm mm} = 5.7 \times 10^{-4}$ every other reflection.

Hence, the total losses {\em per reflection} are $L_{\rm tot} = L_{\rm refl} +
\frac{1}{2} L_{\rm hole} + \frac{1}{2} L_{\rm gaps} = 15 \times 10^{-4}$.
This results in an average of $1/(15 \times 10^{-4}) = 670$ reflections. 

\begin{figure}
\centering\includegraphics[width=0.8\textwidth]{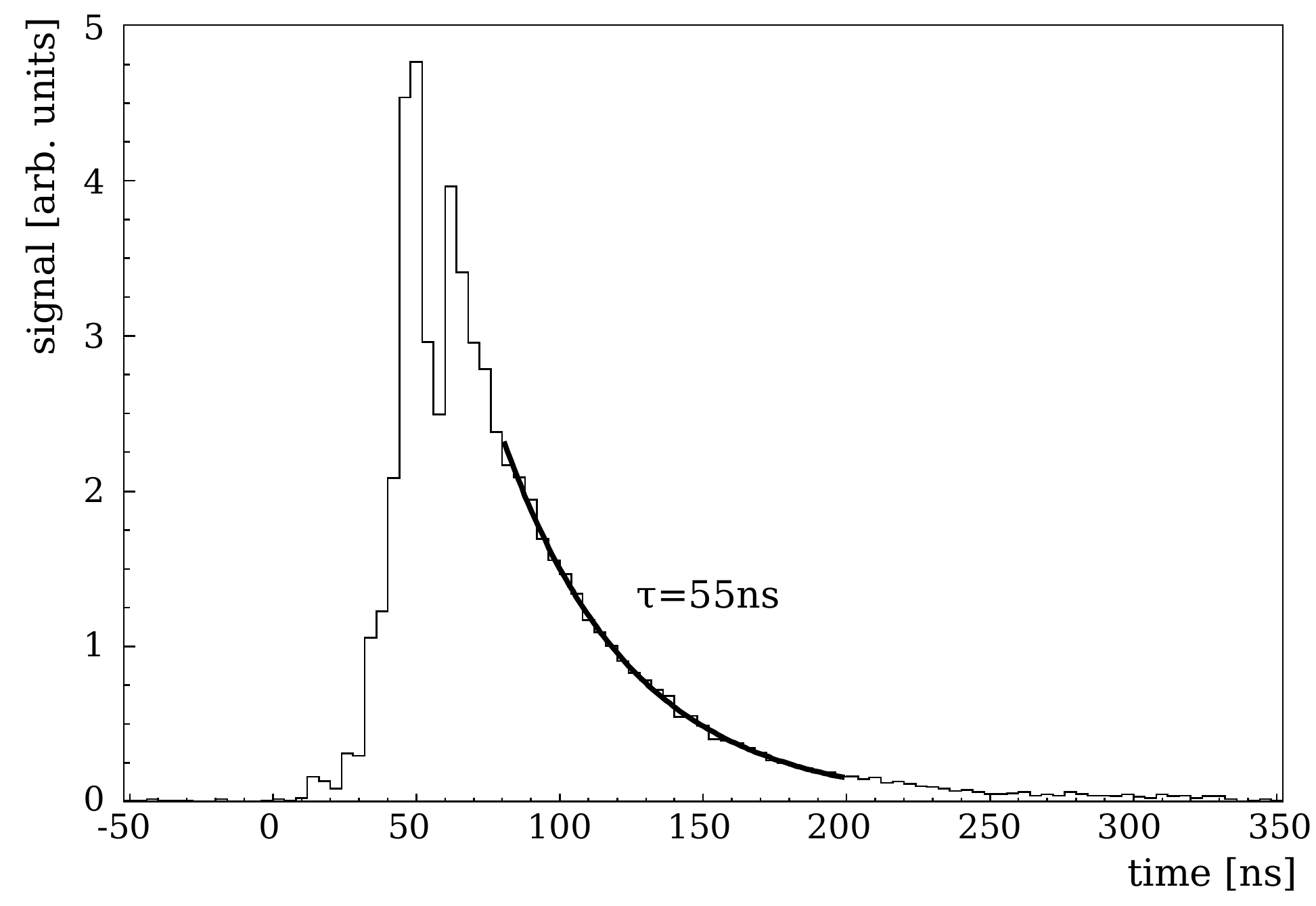}
\caption{Measured time spectrum of the 6\,\mum{} light leaving the
  cavity through the injection hole, recorded 2.5\,m away from the
  mirror Cavity~B.  The structure of the 5\,ns long pulse
  is visible after 1/2 and one round trip (peaks at very early
  times). After 1.5 round trips the light fills the whole cavity and
  decays with a lifetime $\tau_{\rm conf} = 55$\,ns (see
  Sec.~\ref{sec:fluence}).  }
\label{fig:cavity_decay_time}
\end{figure}

With a mirror spacing of 25\,mm (12 reflections per nanosecond) this
should result in a light confinement lifetime of $\tau_{\rm conf} \approx
56$\,ns, in good agreement with the value measured in Cavity~B at
$\lambda = 6$\,\mum{} (Fig.~\ref{fig:cavity_decay_time}).

The pulse energy entering the cavity is 0.15\,mJ.
The resulting average fluence in the illuminated volume with a cross
section of $A_{\rm illumi} = 176 \times 7\,{\rm mm}^2 = 12.3\,{\rm
  cm}^2$ is therefore
$F = 0.15\,{\rm mJ} \cdot 670 / 12.3\,{\rm cm}^2 ~ = ~ 8.0\,{\rm mJ / cm}^2$.

This corresponds to half the saturation fluence of the $2S_{1/2}(F=1)
\rightarrow 2S_{3/2}(F=2)$ transition in muonic hydrogen of 16.5\,mJ/cm$^2$
\cite{Romanov:1995}.
The measurements~\cite{Pohl:2010:Nature_mup1,Antognini:2013:Science_mup2}
confirm this estimate.

\section{Technical considerations}

\label{sec:practical}

\subsection{Mirror \#1}
\begin{figure}[t!]
\centering\includegraphics[width=0.5\textwidth]{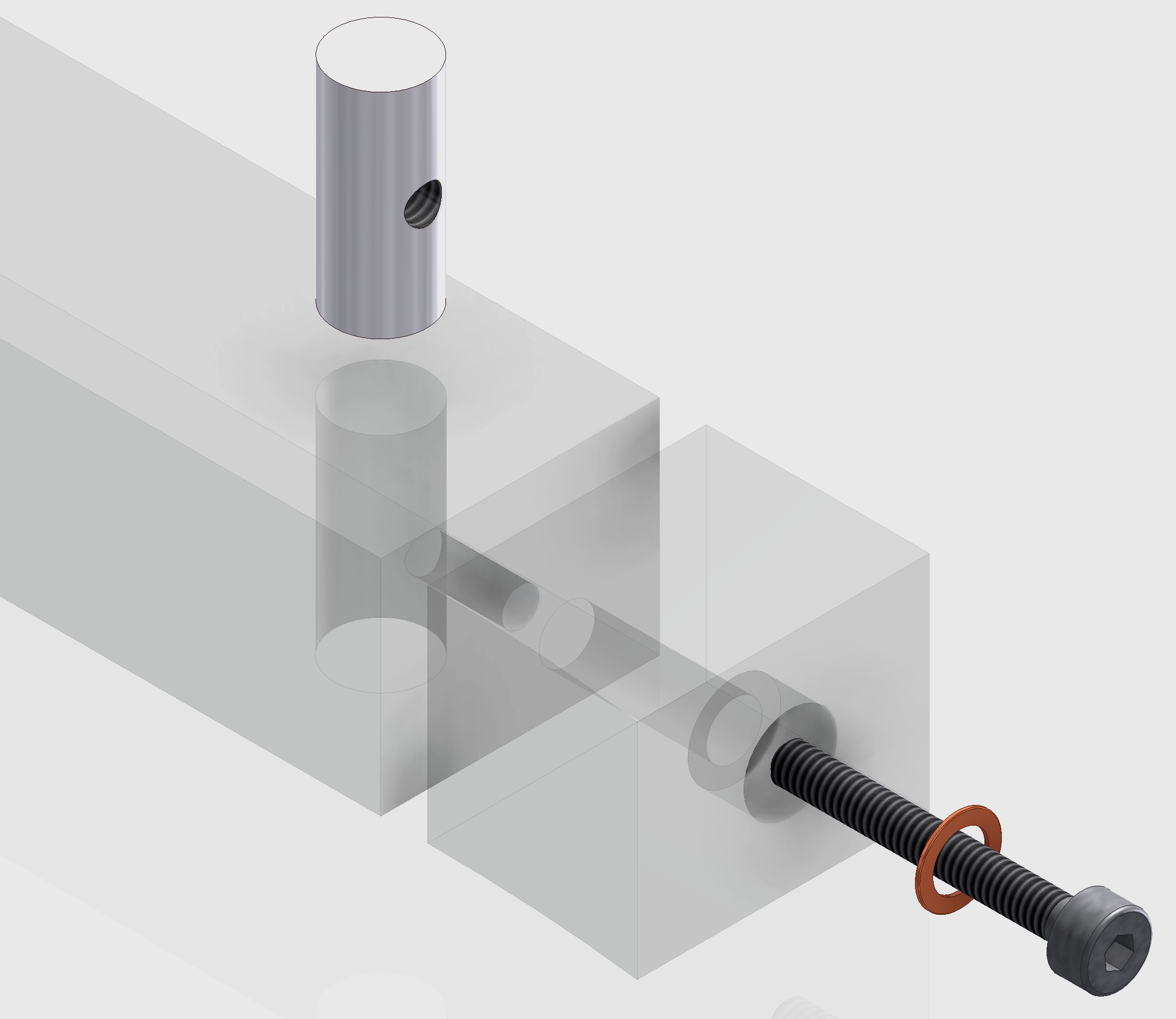}
\caption{Detail drawing of a cylindrical end piece and the parts used
  for mounting it. An aluminum bolt, a copper-beryllium washer and a
  titanium screw are used for fixing the substrate in its final
  position. Compatibility with vacuum, high temperature during
  coating, and a strong magnetic field were reasons for choosing these
  materials.}
\label{fig:explosion}
\end{figure}

The substrates were manufactured by Lens-Optics\cite{Company:LensOptics}.
Mirror \#1 was assembled in our lab. For insensitivity
against the 5\,T magnetic field we used an aluminum bolt, a
copper-beryllium washer and a titanium screw (see
Fig.~\ref{fig:explosion} for details). A microscope was used to adjust
the optical surfaces of the flat and cylindrical parts. Great care
must be taken to avoid rotational misalignment of the small cylinders
with respect to the flat piece. A laser pointer beam was moved across
the gap and the reflected beam was observed a few meters away. We
achieved an accuracy of about 1\,mrad in the ideal case. Such a
precision is mandatory because a misalignment will lead to losses of
the confined light: Light will experience a nonadiabatic vertical
``kick'' on the rotated cylinder which leads to a gradual increase in
the illuminated height. Eventually, light will spill over the upper
and lower edges of the cavity mirrors. Additionally, a horizontal
non-adiabatic kick will be introduced, when the cylinder of the end
piece does not end tangentially with the plain of the main
substrate. Care must be taken during fabrication and when choosing the
end piece.

After aligning the small cylinders with high precision we glued small
glass pieces onto the back side of the mirror, either using Torr
Seal~\cite{Company:Agilent} or EpoTek 353ND~\cite{Company:EpoTek}.  Both
display low outgassing, but the latter can stand higher temperatures
during the coating process.

\subsection{Mirror \#2}
Mirror \#2 is a simple cylinder.
Care must be taken to avoid a toroidal shape of mirror
\#2. Such a torus can easily occur in the manufacturing process of the long
cylindrical mirror. Simulations show that the radius of curvature along the
cylinder axis should be well above 100\,m. Otherwise the light will be
confined to a small central part of the cavity.

\subsection{Coating}
\label{sec:coating}

The dielectric ZnS/Ge coating~\cite{Company:Umicore} of Cavity~B provided a
high reflectivity of $R = 99.89$\,\% over a wavelength range of
$\lambda = 5.5 \dots 6.1$\,\mum{}. This was measured using pulsed
laser cavity ringdown in a 50\,cm long flat-concave cavity built from
1" witness samples.  An even higher reflectivity ($R = 99.97$\,\% or
better) had previously been achieved for Cavity~A by a ThF$_4$/ZnSe
coating~\cite{Company:Lohnstar}, but the small amount of alpha particles emitted from
the coating could not be tolerated in our setup. Cavity~C was recently
used at $\lambda \approx 810$\,nm. The coating~\cite{Company:ATF} has a reflectivity
of 99.97\,\% for $\lambda \in [800 \dots 975]$\,nm needed for the
muonic helium measurements~\cite{Antognini:2011:Conf:PSAS2010}.

\begin{figure}[h]
  \centering\includegraphics[width=0.3\textwidth]{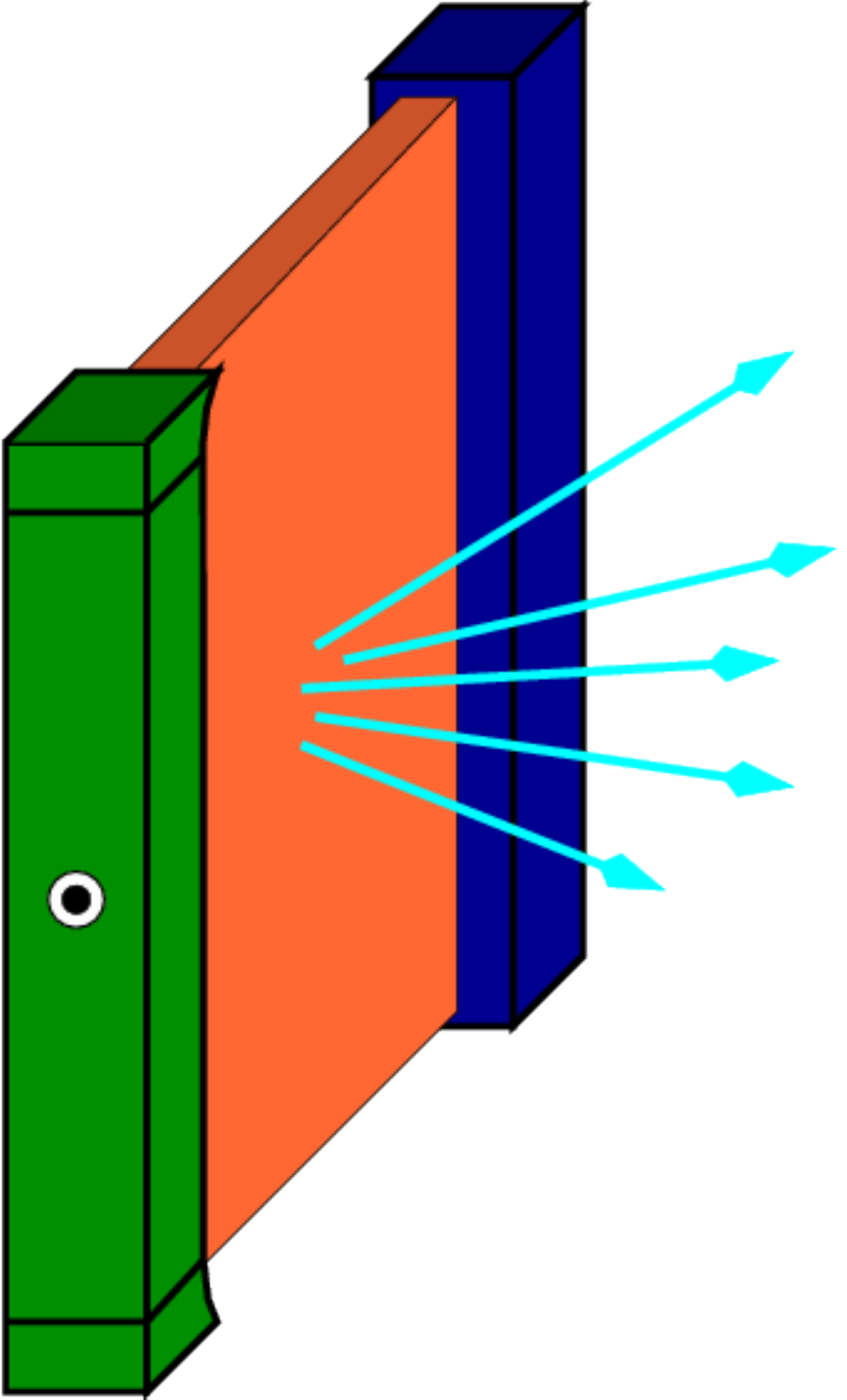}
  \caption{A slightly different geometry of the cavity presented here may be
    used to create a ``light curtain'', for example $20 \times 10 \times
    1$\,cm$^3$ (red)~\cite{Curtain}.
    This can be useful in a variety of experiments, 
    e.g.\ with large or very divergent particle beams (light blue arrows)
    \cite{Antognini:2012:MuoniumEmission,Mills:2014:ultraslow_muons,Crivelli:2011:Conf:PSAS2010,Cassidy:2012:Ryd_Ps,Schott:2009:BoundBeta}
    (see text).}
  \label{fig:curtain}
\end{figure}

\section{Conclusions}

A multipass mirror cavity has been presented that allows illumination of
a very large and very elongated volume. 
Such a volume can not
easily be illuminated by other mirror configurations, as large cavity waists
result in excessive sensitivity to mirror misalignment.
Our cavity achieves an intensity distribution which is homogeneous within
$\pm 30$\,\%, and is very robust against misalignments. No active mirror
stabilization is required.

In the measurement of the Lamb shift in muonic
hydrogen~\cite{Pohl:2010:Nature_mup1,Antognini:2013:Science_mup2} and muonic
helium ions~\cite{Antognini:2011:Conf:PSAS2010} 
such cavities have been successfully used
to illuminate a volume of $176 \times 25 \times 7$\,mm$^3$ 
at wavelengths $\lambda \in [5.5 \dots 6.1]$\,\mum{} and
$\lambda \in [800 \dots 820]$\,nm, respectively. 
Larger volumes can be easily realized.

Such a cavity may be useful for a variety of other experiments. 
Optical excitation of a beam of hydrogen atoms to the metastable 2S state for
a precision measurement of the Rydberg constant~\cite{Flowers:NPL:2007} can
circumvent the systematic uncertainties that originate from electron
bombardment~\cite{Beyer:2013:AnnPhys}.

Another application may be transverse cooling of a beam of atoms or
molecules. Using two such cavities, 2D transverse cooling may be achieved. 
This may for example prove useful for experiments with antihydrogen 
$\overline{\rm H}$ beams for spectroscopy~\cite{ASACUSA:2013:Hbar_beam} 
or the measurement of the free fall of $\overline{\rm H}$ 
atoms~\cite{AEGIS:2011:PSAS,GBAR:2011:Proposal,ALPHA:2013:Hbar_gravity}.

The cavity may also be used to create a ``light curtain'' illuminating a
region of e.g.\ 20\,cm $\times$ 10\,cm, over a distance of a cm or more along
the beam axis~\cite{Curtain}, as sketched in Fig.~\ref{fig:curtain}. 
Laser spectroscopy of Muonium ($\mu^+ e^-$)
\cite{Antognini:2012:MuoniumEmission, Mills:2014:ultraslow_muons} 
or Positronium (Ps $\equiv e^+ e^-)$ \cite{Crivelli:2011:Conf:PSAS2010},
or creation of Rydberg Ps~\cite{Cassidy:2012:Ryd_Ps},
where exotic atoms are emitted 
from a surface and into a large solid angle, could benefit from such a geometry.
Also, optical excitation of H atoms created in neutron
decay~\cite{Schott:2009:BoundBeta} requires transverse illumination of a large
volume inside a neutron guide.

\section*{Acknowledgments}
The authors would like to thank 
K.~Linner, W.~Simon, and the excellent workshops of MPQ and PSI for the support 
in designing and building the various mechanics, and
H.~Br\"uckner of MPQ for help on electronics.
We thank S. Spielmann-J\"aggi and L. Carroll of PSI for measuring substrate and 
mirror properties.
The people at Lens Optics GmbH, dama technologies ag, LohnStar Optics Inc., 
Umicore Coating Services Ltd., and Advanced Thin Films have done a great job.
M.D., J.G., J.K., T.N., and R.P. acknowledge support from the European 
Research Council under StG.\ 279765,
A.A.\ and K.S.\ from SNF 200021L\_138175, and 
T.W.H.\ from the Max-Planck-Society and the Max-Planck-Foundation.

J.V.'s present address is: Institut f\"ur Physik, Carl von Ossietzky 
Universit\"at, 26129 Oldenburg, Germany.
T.W.H.\ is also at: Ludwig-Maximilians-Universit\"at, 80539 Munich, Germany.


\begin{thebibliography}{10}
\bibitem{Curtain}
This particular geometry requires 220\,mm long and 15\,mm high mirrors,
  separated by 100\,mm. The cylinder mirror can have a radius of curvature of R
  = 500\,mm, and the cylindircal end pieces have R = 200\,mm.
\end{thebibliography}
\end{document}